\documentclass[a4paper,10pt]{article}
\usepackage{graphicx}

\begin{document}


\title{On the origin and acceleration of ultra high energy cosmic rays: Cooling flow clusters and AGN hosts\footnote{Published in NIMA, doi:2010-12-018}}

\author{Gizani, Nectaria A. B.\footnote{Hellenic Open University, School of Natural Sciences \& Technology, Physics Laboratory, Tsamadou 13-15 \& Ag. Andreou, 26222 Patra}}
 
\date{}
\maketitle

\paragraph{Abstract}

We are looking for radio `relics' and `halos' in an X-ray selected sample of clusters of galaxies. These radio features are not a product of the Active Galactic Nuclei (AGN)-mechanism, but more likely are associated with past cluster merger events. AGN hosts of cooling flow clusters contain particle bubbles that show non-thermal radio emission. These bubbles could explain the presence of radio relics and halos if they can restrict cosmic rays efficiently. Intracluster magnetic fields and cluster environments can reveal the acceleration mechanisms of cosmic rays.  Using radio/X-ray data and analytical methods we examine three AGN hosts out of our 70 clusters, namely Hercules A, 3C\,310 and 3C\,388. We found that none of these clusters contain relics and/or halos.\\


\section{Introduction}

Ultra-high energy cosmic rays (UHECR) detected by the Pierre Auger and other cosmic ray Observatories suggest that they can be accelerated in the
active galactic nuclei (eg. \cite{dremmer,ensslin}) , in shocks, formed as the jets continuously feed the lobes with new material (eg. \cite{dremmer,no}), and in turbulent magnetized plasma in the jets and/or lobes of radio galaxies. This suggests that super-massive black holes at the centers of FR I and II radiogalaxies can produce high energy particles. Sub-parsec scale acceleration is efficient as long as the scales are comparable to the scale of jet generation or initial collimation.  Hardcastle \cite{hardcastle} suggests that stochastic particle acceleration of UHECRs to high energies (10$^{20}$~eV) is possible within the large-scale lobes of powerful radiogalaxies as long as the radio sources are at low redshift. En$\beta$lin et al., \cite{ensslin96} concluded that radio galaxies are powerful enough to heat and support the cluster gas with injected cosmic-ray protons and magnetic field densities (permitted by Faraday rotation and gamma-ray observations of galaxy clusters) within a cluster radius of $\sim$ 1 Mpc. 

Energy input in the intracluster medium (ICM) heats the gas, and injects cosmic rays (CRs) and magnetic fields. These are the three constituents of ICM, i.e. relativistic particles (CRs), weak magnetic fields (order of $\sim \mu$G) and gas, which is detected in the X-rays. 

Haloes, relics and minihalos are structures of diffuse low surface non-thermal radio emission of relativistic particles spiraling intracluster magnetic fields. They are not directly associated with the AGN phenomenon or the galaxies of the cluster themselves. Their origin is thought to lie in the re-acceleration of relic population of relativistic electrons, or proton proton collisions with the ICM (e.g. \cite{ferrari}).  

We search for these structures in a sample of 70 Abell clusters. The sample was selected upon the radio to X-ray correlation. Our scope is to probe the role that cluster magnetic fields (via Faraday rotation and Inverse Compton Radiation arguments), merger events (through radio/X-ray interactions), cooling flow phenomena, jets, shocks, and nonthermal radio bubbles play in the production, acceleration and propagation of cosmic rays. In the current paper we present the study of three powerful AGN from our sample. These radiogalaxies (RGs) are situated at the center of dense galaxy cluster environments called 'Hercules A', '3C\,388' and '3C\,310'-cluster named after their hosts. 

{\it Hercules A:} 3C\,348, or Her A is optically identified with a cD galaxy at z=0.154, the dimmer and bigger of the two galaxies, at the centre of a poor cluster. It is the fourth brightest radiogalaxy in the sky at low
frequencies. Its radio power at 178 MHz is P$_{178 MHz}^{AGN}= 19\times 10^{26}$ W Hz$^{-1}$ sr$^{-1}$. Although it is a high luminosity source, its structure resembles that of low luminosity Fanaroff-Riley class I objects, no hotspots and brightened edges (typical of FR IIs). These characteristics classify the radio source as an FR1/2. Its unusual jet dominated morphology presents interesting differences: helical structures in the eastern jet and striking ring-like features in the western one \cite{dreher}. HST data suggest possible kiloparsec rings of obscuration, aligned near the radio axis with a slight offset from the galaxy nucleus \cite{baum}. 

{\it 3C\,388:}  The optical galaxy (z = 0.0908) is one of the brightest cDs. In the radio is a relatively small classical double FRII. Its radio power is P$_{178 MHz}^{AGN}= .48\times 10^{26}$ W Hz$^{-1}$ sr$^{-1}$ at 178 MHz. 
Its faint halo in the eastern lobe with steep spectrum and high polarization, and the relic of an older jet activity are associated with the individual galaxy rather than the cluster as a whole. The western side presents a similar low surface radio emission. There is a luminous jet contained within  the western lobe and a possible counter-jet \cite{rot}.

{\it 3C\,310:}  3C\,310 is identified with a cD galaxy at z=0.054. Its radio power at 178 MHz is P$_{178 MHz}^{AGN}= .37\times 10^{26}$ W Hz$^{-1}$ sr$^{-1}$ \cite{van}. HST data revealed an optical\cite{martel}. The X-ray bolometric luminosity of the cluster is L$_{X}^{bol}= 10^{36}$ W, while the X-ray luminosity associated with the AGN-host is L$_{X}^{AGN} \sim 7\times 10^{35}$ W. The X-ray emission is found to be elongated perpendicular to the radio jet axis \cite{miller}. 

For this work we assume H$_{\circ}$ = 65 km s$^{-1}$ Mpc$^{-1}$ and q$_{\circ}$ = 0 throughout.

\section{Observations and Results}

\subsection{Hercules A}

We have studied Her A in detail, both in kpc- and pc-scales. We have used VLA multiconfiguration, multifrequency interferometry to take total intensity and polarization data at 1.4 arcsec resolution \cite{gizanir}. EVN observations focused on the rather weak radio core for such a bright source \cite{gizanie}. 
In kpc-scales we have found a dramatic spectral asymmetry between the two sides of the radio emission: The jets and rings have a flatter spectrum than the surrounding lobes and bridge, strongly suggesting that we are whitnessing a recently renewed outburst from the active nucleus . The lobes have steeper spectrum than found in typical radio sources, and steepen further towards the centre. We interprete this asymmetry between the diffuse lobes in terms of relativistic beaming together with front-back travel delays. This actually means that we are  viewing the two lobes at different stages of the outburst. 

The compact core of Her A is optically thin and presents a remarkably steep spectrum. It is still unresolved and very weak ($\sim$ 15 mJy) even at 18 miliarcsec (mas) EVN resolution. The kpc-jets although collimated at short distances from the core within the lobes, they are misaligned in pc-scales with a $ \approx 35^{\circ}$ angle of misalignement in NW/SE direction, which could be suggesting a dense environment. 

The Hercules A cluster is luminous in Xrays, although poor in optical galaxies.  Its X-ray bolometric luminosity is L$_{X}^{bol}=48\times 10^{36}$ W, while the X-ray luminosity associated with the AGN is L$_{X}^{AGN} = 20\times 10^{35}$ W.
Our ROSAT PSPC and HRI X-ray observations \cite{gizanix} revealed a dense environment, in which Her A lies as the host, typical for modest cooling flows. The PSPC spectrum shows a cool component of the ICM with 0.5$\leq$ KT(keV) $\leq$1. The result added to the BeppoSAX and ASCA findings suggest  a multiphase gas. The cluster emission is elongated parallel to the radio axis extending further than the radio lobes. Although our low sensitivity data found no obvious displacement of the ICM by the relativistic plasma, Chandra observations have confirmed confinement of the inner jets by the ICM and the presence of cavities \cite{nulsen}. These cavities suggest depressions of the X-ray emission, coincident with the radio lobes, as the jets push the gas. However Nulsen et al., 2008 suggest that the X-ray cavities and the front shock (cocoon) observed, are not associated with the radio lobes which is rather unusual.  

Our X-ray analysis (central electron density) together with the radio one (rotation measure) allowed the study of the environment of the AGN in terms of its external magnetic field. We have estimated that the magnetic field in the ICM has a radial dependence of the form  B(r)$\propto n(r)^{m-1}$, where $n(r)$ the electron density given by $n(r) = \frac{n_e\circ} {[(r/r_{\circ})^{2}+1]^{3\beta/2}}$. n$_{e\circ}$ is the central value of the electron density and the parameter $\beta $ is obtained by a modified King model fitting of the azimuthally averaged X-ray surface brightness profile. The tangling scale size is found to be 4$\leq$ D$_\circ$ (kpc) $\leq$ 35. 

\subsection{3C\,388}

3C\,388 is at the centre of a poor cluster with very dense ICM. It is likely to be a cooling flow cluster. Its X-ray bolometric luminosity is L$_{X}^{bol}=5\times 10^{36}$ W. 
Our ROSAT HRI observations (energy range 0.1--2.4 keV) have found warm gas confining the lobes of this powerful radio source. The X-ray luminosity associated with the AGN is L$_{X}^{AGN} = 5\times 10^{35}$ W. Emission is detected to a radius $\simeq 5 \times$ larger than the radio lobes. We found marginal evidence for distortion of the gas by the radio lobes.  Chandra data cleared this ambiguity \cite{kraft}. The analysis of the new data gave $\beta = 0.444 \pm 0.003$, central hydrogen density of 8.3 $\cdot 10^{4} m^{-3}$, gas temperature kT $\simeq$ 3.5 keV (although varying slightly between the nucleus and lobes region) and r$_c$=15 kpc.

\subsection{3C\,310}

We study this source as it presents many similarities to Hercules A: They both have double optical nuclei of similar absolute magnitude in R-band at low redshift. They are 
 at the center of cooling flow clusters of galaxies with similar gas temperature and a contribution from a point source. They are classified as FR1.5 and have sharply bounded double lobes. Their lobes present asymmetry with respect to brightness, depolarisation and spectral index. They have no compact hotspots. Instead they are probably the only two sources which contain such clear ring-like radio features. Other high-brightness structure is also present with flatter 
spectra than the surrounding diffuse lobes suggesting a periodically active nucleus. Their projected B-field follows closely the edges of the rings and lobes. They are both old sources with a steep spectral index$\alpha \simeq$ -1.4 (assuming that the flux density is given by S$_\nu \propto \nu^{\alpha}$). Their thermal pressure at the distance of the radio lobes is greater than the lobe minimum pressure. 

We have made global VLBI observations at 18 cm, 4 mas resolution to map the source's core. The core is unresolved at kpc-scales and has a flux of $\sim$ 130 mJy at 21~cm, at 4 arcesec. We have found two compact components at the core region at 5 pc-scales. We have detected the 7.3\% of the VLA flux.   The position of the core is still unknown. The data seem to reveal a north-west (NW) emission maybe associated with the pc-scale jet. If this is true then there is a misalignment of 20$^{\circ}$ with respect to the NW kpc-scale jet. Another emission is detected in NE/SW direction (misaligned $\sim 100^{\circ}$ w.r.t. the NW direction). We suggest that the pc-scale asymmetry is due to Doppler beaming: the mas-jet is always on the side of the brightest kpc-jet with respect to the core. The misalignment between the pc-, kpc-jets is a new similarity between Her A and 3C\,310 revealed by our analysis.

Table~\ref{all} summarizes the results of the model fitting to the gas distribution of the observed clusters. 

\begin{table}
\begin{center}
         \begin{tabular}{llllll}
            \hline
            \noalign{\smallskip}
             Cluster   &\,\,\,\,\,\,\,\, $\beta$  & n$_{e\circ}$ & kT & r$_c$ & r$_{tot}$ \\
           &  & 10$^4$ m$^{-3}$  & keV & kpc  & kpc \\ 
           \noalign{\smallskip}
            \hline
            \noalign{\smallskip}
             Her  \,\,A & 0.74 $\pm$ .03 & 1 & 0.5-1 & 121 & 2200 \\
           3C\,388 &  0.53 $\pm$ .04 &  1.5  & 3 & 33 & 180 \\
           3C\,310 & 0.5 &  0.2  & 2.5 & 84 & $>$ 670 \\

            \hline
         \end{tabular}
\caption{Parameters from the model fitted to the gas distribution. From left to right: Cluster-AGN id; $\beta$ parameter ; central electron density n$_{e\circ}$; temperature of the gas; core radius r$_c$; total X-ray extent r$_{tot}$. \label{all}}
\end{center}
 \end{table}

\section{Magnetic fields, Cosmic rays, AGN Energy budjet}

Mergers and AGN hosts in clusters inject and accelerate magnetised plasma, i.e. magnetic fields and energetic particles filling the ICM. This continues during the lifetime of clusters. The combined energy of CRs and fields should be comparable to the present-epoch thermal energy content of the central region. However the field topology is expected to be complicated, with local variations of the field strength and turbulence present (eg. merger events, jet-phenomenon, etc). All these suggest an unstable and time-dependent configuration of fields, cosmic-rays and thermal gas. The presence of turbulent magnetic fields in the ICM of the cluster as well as merger activity most likely show up as radio halos, relics and can accelerate cosmic rays. 

Her A, 3C\,388 and 3C\,310 are cD galaxies and powerful AGN at the centre of cooling flow, dense, but poor clusters. cD galaxies and powerful radiosources could be produced through highly anisotropic mergers \cite{west}. 
We have found  no radio haloes, relics present which would be expected as a result of a merger event. X-ray observations of 3C\,388 revealed that there could be a possible sub-cluster either falling towards the host or already passed through \cite{kraft}. The existing radio data show no radio structure associated with that. 

However distortion of the hot X-ray emitting intracluster gas in the form of holes (cavities) is expected, as the radio jets push their way through the gas as they travel in the ICM.  Cavities are found in the 3C\,388 cluster coincident with the lobes. A Chandra proposal for detecting the expected cavities in the 3C\,310 cluster has currently been submitted. The radio/X-ray correlation could either suggest that the relativistic electrons are distributed homogeneously over the lobe, whereas the magnetic field is amplified towards the rim of the lobe region.  
The fact that the X-ray cavities discovered in Her A not supposed to be associated with the source's radio lobes, contrary to what it was expected, could suggest that the evolution of the radio-galaxy cannot be determined entirely by the global properties of the hot gas.  

Similarly to the trapping of CRs in the frozen-in magnetic field to the ionized interstellar gas in the Galactic plane, we can assume that CRs could also be trapped by the intracluster magnetic field. We have estimated the thermal energy content of the two clusters by using a $\beta$-model for the cluster gas. Results of the magnetic field estimates are shown in Table~\ref{all2}. The magnetic field for 3C\,310, assuming equipartition, is estimated to be 5$\mu$G \cite{van}. We mention here that intracluster magnetic fields from Faraday screen observations for the powerful host RGs of the Hydra A and Cygnus A clusters revealed values as high as $\sim 30 \mu$ G. Central estimates of cluster magnetic fields of the order of 10 $\mu$ G could suggest a nonthermal phase. Such strengths are directly observed in clusters with central radio galaxy and cooling flow, and could probably be typical.

However there is another phase which consists of cosmic ray protons, that have cooling times equal to or larger than the Hubble-time. The radiative cooling time (in Gyr) in the central regions of the Her A cluster is $\simeq$ 6 Gyrs for the hot phase and 2 Gyrs for the cool phase. For the  3C\,388 cluster is 3.7 Gyrs. Comparison of these values with the age of the universe ($\sim$ 10~Gyrs) implies the existence of a cooling flow suggested by the X-ray data \cite{sarazin}. All clusters studied here are cooling flow clusters. They also present extended radio emission. Synchrotron cooling time is too short for the relativistic electrons to diffuse in the whole lobes considering the growth speed of the lobes \cite{scheuer}. The short cooling time of the emitting cosmic ray electrons and the large extent of the radio sources suggest an ongoing acceleration mechanism in ICM. 

While the radio emitting medium leaves the lobes diffusing into the cluster medium, the CR part of the outflow is dissipated. Relativistic electrons loose their energy via numerous
cooling mechanisms: relatively quickly through synchrotron
emission whilst spiraling the cluster magnetic fields ; also through Compton scattering through collisions with photons of the cosmic microwave background. These processes are not effective for energetic protons whose Compton and synchrotron cooling times are much longer
than the Hubble-time. A possible mechanism for slowing down energetic protons is suggested via electronic excitations (see Ensslin et al. \cite{ensslin96} and references therein). 

Cooling of the electrons is 
visible in the steepening of the spectral index of radio emission found close to studied radio galaxies in the clusters. Steep spectral indices found in the lobes of Hercules A, 3C\,388 and 3C\,310 imply short lifetimes of radiating particles and also re-acceleration of the electrons to some extent. This would result in energy redistribution in the ICM. If cooling is significant to the Hubble time then the cooling radius of the Her A cluster is $\sim$ 90~kpc less than the core radius of $\sim$ 120~kpc found by our X-ray analysis. The energy lost by the energetic particles could be gained by the magnetic fields also heating the ICM. The clusters seem to grow. As a result the energy of CRs in the ICM should increase adiabatically. Additional CR sources such as supernovae, shocks, in-situ acceleration could also compensate for the energy loss. 

Our radio analysis (after correcting for Faraday rotation) showed that the projected magnetic field closely follows the edges of the lobes, the jets and the rings of Her A. Our X-ray analysis \cite{gizanix,leahy} has pointed out that the gas thermal pressure is greater than the minimum pressure in the radio lobes by an order of magnitude for our best estimate. This implies confinement of the radio structure of the AGN by the ambient ICM rather than by shocks. The same is true for both 3C\,388 and 3C\,310. Confinement indicates that the  interior should be dominated by particle pressure, while magnetic pressure should dominate in the shell region defined by the lobe boundary.  There is little entrainment, so the energy supply of the lobes mostly comes from relativistic particles and magnetic fields. To retain equipartition our results would require B $\approx$3B$_{me}$, where B$_{me}$ is the magnetic field found using minimum energy. However as discussed by Leahy \& Gizani \cite{leahy}, the magnetic field should be below equipartition and therefore unimportant to the lobe dynamics. "Invisible" particles (relativistic protons, low energy  e$^{-}$ / e$^{+}$) should be the dominant particles in the jets, lobes. This is also confirmed by Ensslin et al. \cite{ensslin96} who suggested the production of
very energetic relativistic protons, besides relativistic electrons, in the lobes of radio galaxies to explain cosmic ray energies $> 3 \cdot 10^{18}$ eV reaching Earth. Jets with particle composition of low energy e$^{-}$/e$^{+}$ and relativistic protons \cite{leahy,leahya} could be energetic sources of cosmic rays. Similar results have also been suggested theoretically \cite{ensslin}. 


Following the analytical model fitting by En$\beta$lin et al. \cite{ensslin96} of the correlation between the RGs' jet power vs luminosity at 2.7 GHz, we estimate the energy input into
the central region of clusters from the host RG. This yields to $\simeq 1.7 \cdot 10^{22} W kpc^{-3}$ assuming a power law with index b=0.82. The injected jet power may dissipate and heat the gas, or could accumulate and support the ICM (magnetic fields and particles). We have calculated the central cosmic ray energy $\epsilon_{CR(r)}$ by using 
$\epsilon_{CR(r)} = 3n_{\circ e}(r)\cdot kT\cdot \alpha_{CR}$  as the thermal energy density, assuming the scaling ratio between the thermal and CR energy densities to be $\alpha_{CR} \simeq 1$. 

The production rate of gamma rays above 100 MeV by $\pi_{\circ}$-decay after hadronic interactions of the energetic protons with the background gas is estimated using  
$dn_{\gamma}(> 100MeV)/dt = 3 q_{\gamma} n_{e\circ}^{2} kT \alpha_{CR}$ . The parameter $q_{\gamma} = 0.39 \cdot  10^{-12} m^{3} W^{-1} s^{-2}$ applies to a proton
spectrum similar in slope to that observed in the Galaxy. See Table~\ref{all2} for the results.

\begin{table}
\begin{center}
        \begin{tabular}{lllll}
            \hline
            \noalign{\smallskip}
            Cluster   &  B$_{RM}$  & B$_{IC}$ & \,\,\,\,\,\,\,\,\,\,\,\,$\epsilon_{CR}$ & \,\,\,\,\,\,\,\,dn$_{\gamma}$/dt \\
           &   $\mu$ G & $\mu$ G & 10$^{-12}$ m$^{-3}$ Ws$^{-1}$ & 10$^{-20}$ m$^{-3}$ s$^{-3}$ \\
            \noalign{\smallskip}
            \hline
            \noalign{\smallskip}
            Her  \,\,\,A &  3-9 & \,\,\,\,4.3 &\,\,\,\,\,\,\,\,\,\, 2.4-4.8 & \,\,\,\,0.94-1.9\\
            3C\,388 &  \,\,\,\,\,\,-& \,\,\,\, 3.8&  \,\,\,\,\,\,\,\,\,\,\,\,\,\,\,\,21.6 &  \,\,\,\,\,\,\,\,12.6 \\
            3C\,310 & \,\,\,\,\,\,-& \,\,\,\, 3.6&  \,\,\,\,\,\,\,\,\,\,\,\,\,\,\,\,2.4 &  \,\,\,\,\,\,\,\,0.19 \\

            \hline
         \end{tabular}

\caption{Cluster Parameters. From left to right: Cluster id; central value of the external magnetic field using Faraday Rotation; center value of the cluster  magnetic field derived using inverse compton arguments; central cosmic ray energy $\epsilon_{CR}$; production rate of gamma rays dn$_\gamma$/dt, see text.}
\label{all2}
\end{center}
\end{table}

The 408 MHz-luminosity versus size diagram by \cite{hardcastle}, sets a limit on the luminosity and size that a radio lobe of a RG should have in order to be able to marginally confine a particle of energy E$_p$ and charge Z. The adopted model assumes spherical lobes and stochastic particle acceleration within them. The diagram suggests that only reasonably luminous radio galaxies can accelerate UHECR to energies above 10$^{20}$ eV.  The linear size in kpc (largest extent) of Her A, 3C\,388 and 3C\,310 is $\sim$ 540, 92, and 342 respectively, suggesting that all 3 powerful radio galaxies are capable of confining UHECR with energies of 10$^{20}$ eV. 

\section{Conclusions}

In the current work we have confirmed the presence of extragalactic magnetic field in the ICM capable of accelerating cosmic rays. There are no radio halos/relics associated with the studied clusters but depressions in the X-ray emission coincident with the radio lobes. Our observations suggest that the jets of the host AGN should contain low energy e$^{-}$, e$^{+}$, and relativistic p and are capable of feeding cosmic rays. All three radio galaxies are able to  accelerate ultra high energy cosmic rays with energies up to 10$^{20}$ eV.

\bibliographystyle{elsarticle-num}

\end{document}